\documentclass[aps, pra, reprint, longbibliography, 
superscriptaddress]{revtex4-1}

\usepackage{hyperref}
\usepackage{graphicx}
\usepackage[utf8]{inputenc}
\usepackage{xcolor}
\usepackage{amsmath, amssymb}
\usepackage{physics}

\begin{document}

\title{ Supersolid Stripes Enhanced by Correlations in a Raman 
Spin-Orbit-Coupled System}

\author{J. S\'anchez-Baena}\email{juan.sanchez.baena@upc.edu}
 \author{J. Boronat}\email{jordi.boronat@upc.edu}
 \author{F. Mazzanti}\email{ferran.mazzanti@upc.edu}%
\affiliation{%
 Departament de F\'isica, Universitat Polit\`ecnica de 
Catalunya,
Campus Nord B4-B5, E-08034, Barcelona, Spain\\
}%

\date{\today}

\begin{abstract}
A Bose gas under the effect of Raman Spin-Orbit Coupling (SOC) is
analyzed using the Discrete Spin T-moves Diffusion Monte Carlo
method~\cite{sanchez, mitas}. By computing the energy as well as the
static structure factor and the superfluid fraction of the system, the
emergence of an energetically favorable supersolid stripe state is
observed, which is in agreement with recent observations.
A significant enhancement of the stability of the stripe phase with
respect to the mean-field prediction is observed when the strength of
the inter-atomic correlations is increased.
We also quantify and characterize the degree of superfluidity of the
stripes and show that this quantity is mostly determined by the ratio between the Raman
coupling and the square of the momentum difference between the pair of SOC inducing laser beams.
\end{abstract}

\maketitle

\section{\label{sec:introduction}INTRODUCTION }

Spin-Orbit Coupling (SOC), which denotes the interplay between a
particle's momentum and its spin, has been a subject of interest in
the recent years, both theoretically and experimentally. This is due
to the wide variety of exotic quantum states induced by this kind of
interaction, which include topological insulators~\cite{hasan},
topological superconductors~\cite{sato}, and Majorana
fermions~\cite{wilczek}. SOC is a relativistic effect that emerges
naturally in electronic systems, and that is also synthetically
engineered~\cite{review} in ultracold atomic gases.  These recent
realizations in dilute gases represent an important achievement in the
study of the physics of SOC due to the high controllability and
tunability of these systems. In the particular case of Raman SOC, its
implementation was first achieved experimentally by inducing a Raman
coupling via two laser beams on an atomic Lambda type
configuration. SOC is then generated by the simultaneous driving of a
spin flip transition and transferring of
momentum~\cite{spielman,zhang19,wang,cheuk,liu}.  Under this scheme,
Raman SOC has been realized with $^{87}$Rb bosons, both in the
continuum~\cite{spielman} and in a lattice~\cite{engels1,engels2},
and also with other species: $^6$Li~\cite{zwerlein}, $^{40}$K~\cite{zhang2},
$^{87}$Sr~\cite{ye}, $^{173}$Yb~\cite{jo,fallani}, and
$^{161}$Dy~\cite{lev}. In this context, two hyperfine states of the
atom are labeled as the spin states.

In this paper, we focus on Raman
SOC, which couples the linear momentum of an atom with its spin
according to
\begin{equation}
 \hat{W}^{\text{SOC}} =
\frac{ \hbar k_0}{m} \hat{P}_x \hat{\sigma}_z +
\frac{\hbar^2 k_0^2}{2m} - \frac{\Omega}{2}
\hat{\sigma}_x   \ , 
\label{Wraman}
\end{equation}
with $m$ the mass of the particle, $\hat{P}_x$ the $x$-component of
the momentum, $\hat\sigma_x$ and $\hat\sigma_z$ 
the Pauli matrices, $\Omega$ the
Raman coupling, and $k_0$ the magnitude of the wave vector
difference between the two laser beams.
Some striking features induced by the SOC interaction can be observed
already at the single particle level. The coupling between momentum
and spin implies that the minimum of the energy dispersion relation is
in a non-zero momentum, degenerate state for a given range of values
of the Raman coupling~\cite{stringari}. This degeneracy involves
states of equal magnitude but opposite sign in momentum space,
enabling the possibility of a stripe phase ground state. The inclusion of interactions changes this behavior, and depending on the parameters of the Hamiltonian a single momentum state or a stripe state is favored, each one with different momentum.



Supersolid stripes
arise from the breaking of two symmetries: a gauge symmetry, giving
rise to off-diagonal long-range order, and spatial symmetry, seen as a
periodic density modulation in space~\cite{stringari_stripes}.  The
emergence and characterization of stripes in SOC systems have been a
subject of major relevance in the field, both from the
theoretical~\cite{liu2,stringari_stripes} and experimental
sides~\cite{ketterle}. Despite being predicted by theory, the stripe
phase was not detected in the first experimental realization of Raman
SOC by Spielman's group~\cite{spielman}, mainly due to the
extremely low spin dependence of the inter-atomic interactions between
$^{87}$Rb atoms. Later on, Ketterle's group~\cite{ketterle} provided evidence of its
existence through Bragg scattering,
using a new Raman SOC setup with $^{23}$Na that
allowed for a better control of the spin interactions. Recently, they have
also been detected in another experiment with $^{87}$Rb atoms
~\cite{spielman2}, where the contrast of the stripes
is enhanced by rapidly increasing the Raman coupling
before probing the density
modulations by optical Bragg scattering.

In previous theoretical works, the phase diagram of an interacting
system of atoms under Raman SOC has been
reported~\cite{stringari,stringari_2}. However, the diagram is
restricted to the mean-field regime, valid only for very low gas
parameter values $\leq 10^{-4}$.
In order to extend these results to the
  stronger interacting regime,
and to deal with the non-local character
of the SOC interaction, we use the Discrete Spin T-moves Diffusion
Monte Carlo method (DTDMC)~\cite{sanchez,mitas,tmoves} to study the
system from a microscopic point of view. Starting from a variational
ansatz, we propagate the initial wave function in imaginary time
keeping its phase constant, leading to a statistical representation of
the best possible wave function given a phase constraint (fixed-phase
approximation). DTDMC is then used to sample relevant observables,
some of which may not be easily calculated at the mean-field level.
An exact form of the
imaginary time propagator up to first order in the time step is
employed.

This paper is organized as follows. In
  Sec.~\ref{sec:hamiltonian} we discuss the Hamiltonian of the system
  and the relevant parameters used in the elaboration of the
  phase diagram. In Sec.~\ref{sec:phase_diagram} we show and discuss
  the phase diagram of the system, as well as the static structure
  factor, the pair distribution function and the superfluid fraction,
  focusing on the stripe phase. Finally, in Sec.~\ref{sec:conclusions}
  we summarize the main conclusions of our work.


\section{\label{sec:hamiltonian}HAMILTONIAN }

We study a three-dimensional system of $N$ bosons of mass 
$m$ under periodic boundary conditions (PBC) described by the Hamiltonian
\begin{equation}
\hat{H} = \sum_{i} \left[ \frac{\hat P_i^2}{2m} + \hat{W}_i^{\text{SOC}} \right] +
\sum_{i<j} \hat{V}_{ij} \ ,
\label{Hamiltonian}
\end{equation}
where $\hat{V}_{ij}$ is a short-range, two-body, spin-dependent
interaction. We use two model interactions: a soft-sphere (SS)
potential of strength $V_0(s_i,s_j)$ and range $R_0(s_i,s_j)$, and a
Lennard-Jones (LJ) force $V_{ij}(r_{ij},s_i,s_j) = \left( \frac{
  \sigma_{12}(s_i,s_j) }{r_{ij}} \right)^{12} - \left( \frac{
  \sigma_{6}(s_i,s_j) }{r_{ij}} \right)^{6}$. Here, $r_{ij}$ is the
distance between the $i$-th and $j$-th particles and $s_i,\text{} s_j
= \pm 1$ are their spins.  The trial wave function used for importance sampling in the DTDMC method, that also fixes the phase, is chosen as a product of one-body and two-body (Jastrow) terms.  For the former,
we use the expression reported in Ref.~\cite{stringari}, with the sign
of the spin-down component changed due to the different sign of the
$\Omega$ term in the Hamiltonian. The Jastrow factor depends on the
interaction $\hat V_{ij}$. For the SS potentials we use the
zero-energy solution of the averaged interaction along the different
spin channels, which provides a lower variational energy than a spin-dependent two-body Jastrow factor. In the case of the LJ interaction a McMillan factor of
the form $e^{-(b/r)^5}$ is used, with $b$ a constant that is
variationally optimized.  The choice of the parameters of the two-body
interaction $\hat V_{ij}$ determines the different channel scattering
lengths $a_{s,s'}$, as according to Ref.~\cite{raman_scattering} the
inclusion or not of the SOC term does not appreciably change them.
The values used in this work fulfill the condition $a_{+1,+1} =
a_{-1,-1} > a_{+1,-1} = a_{-1,+1}$, as in the experiments of
Ref.~\cite{ketterle}.  Finally, we express all quantities in
dimensionless form, introducing characteristic length ($a_0 =1 /k_0 $)
and energy ($E_0 = \hbar^2 k_0^2 / 2 m $) scales.

In order to characterize the phase diagram of the model, we use the
standard gas parameter $n a^3$, with $a=a_{+1,+1}$ the scattering
length of the interaction in the $(+1,+1)$ channel. It should be
noted that, for this system, $n a^3$ is not a scaling
parameter. However, we use it to characterize the combined effect of
the density and the interaction. We have checked, though, that for
very low values of $n a^3$ one recovers the mean-field results,
while for larger values, 
the DTDMC simulations reveal
that the extension of the stripe phase domain is increased with
respect to the mean-field prediction. This may be a relevant issue for
experiments willing to detect and/or characterize the stripe phase.
In order to illustrate that, we set the density to $n=3.7 \!\cdot\!
10^{-3}$, with the number of particles $N\in [50, 120]$ and the size
of the simulation box changing as a function of the momentum of the
trial wave function. We tune the
spin-dependent scattering lengths such that $n a^3 \in (
10^{-4},\text{} 10^{-1} )$ by changing the two-body potential parameters. In this sense, increasing the gas parameter is equivalent to increasing the range and
strength of the interactions, which enhances the effect of correlations in the medium.
We set the interaction contrast $\gamma = ( a - a_{+1,-1} )/( a +
a_{+1,-1} )$ to $\gamma=0.4$, since non-zero values of this quantity are necessary for the existence of a stripe ground state~\cite{stringari}. The quantitative characterization of the superfluidity in the
stripe phase is performed with the contrast used in Ref.~\cite{ketterle}, 
$\gamma = 0.904$.
It must be remarked that the quantity $\gamma$
is a tunable property in the experimental setup of
Ref.~\cite{ketterle}.

\begin{figure}[t]
\centering
\includegraphics[width=0.9\linewidth]{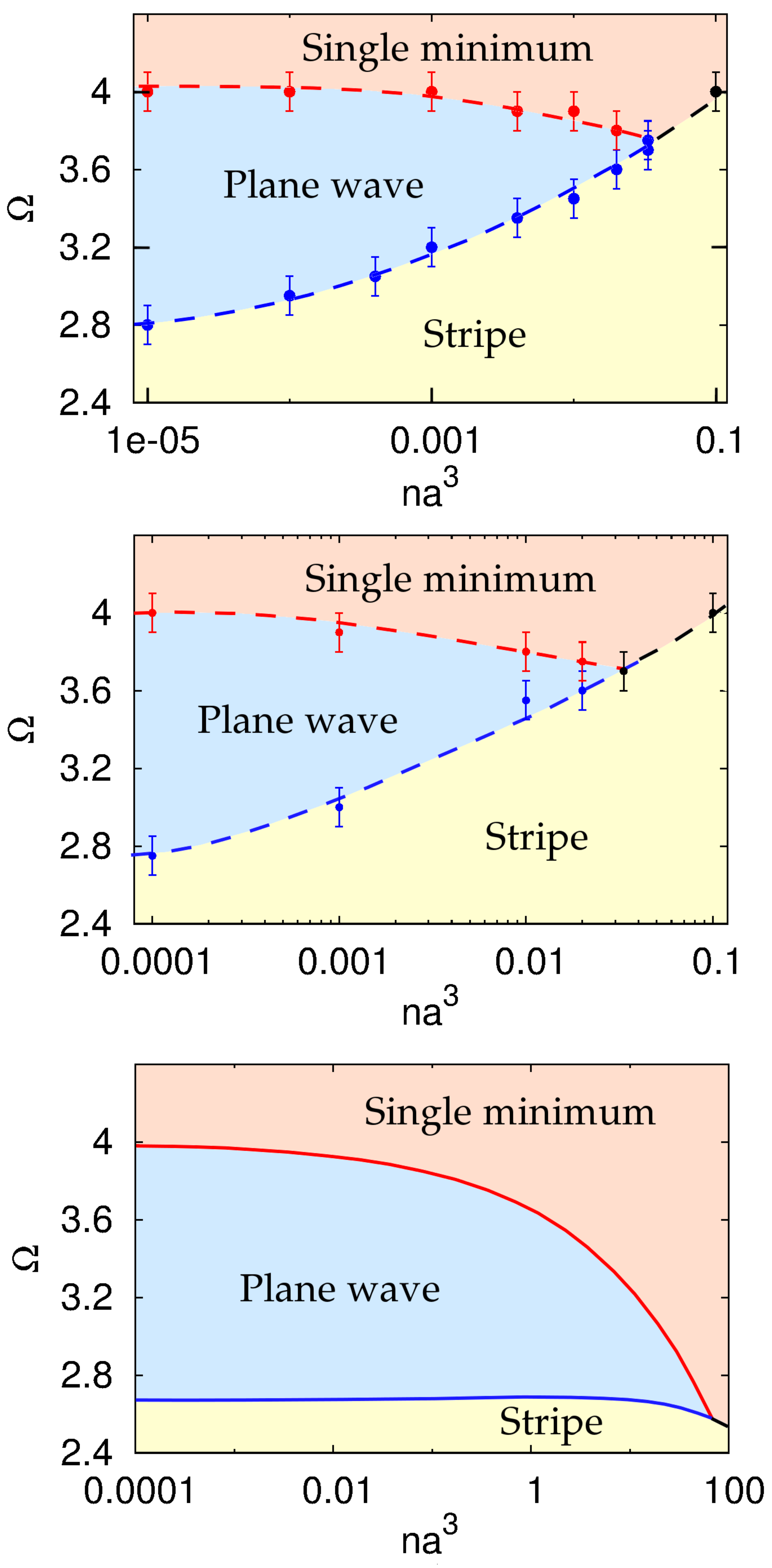}
\caption{Phase diagram of the many-body system with Raman Spin Orbit 
Coupling. The upper plot corresponds to the DTDMC diagram using the SS  
potential and  the middle one to the LJ potential. In 
the lower plot, we report  the mean-field phase diagram.} 
\label{fig_diagrams}
\end{figure}  

\begin{figure}[b]
\centering
\includegraphics[width=0.85\linewidth]{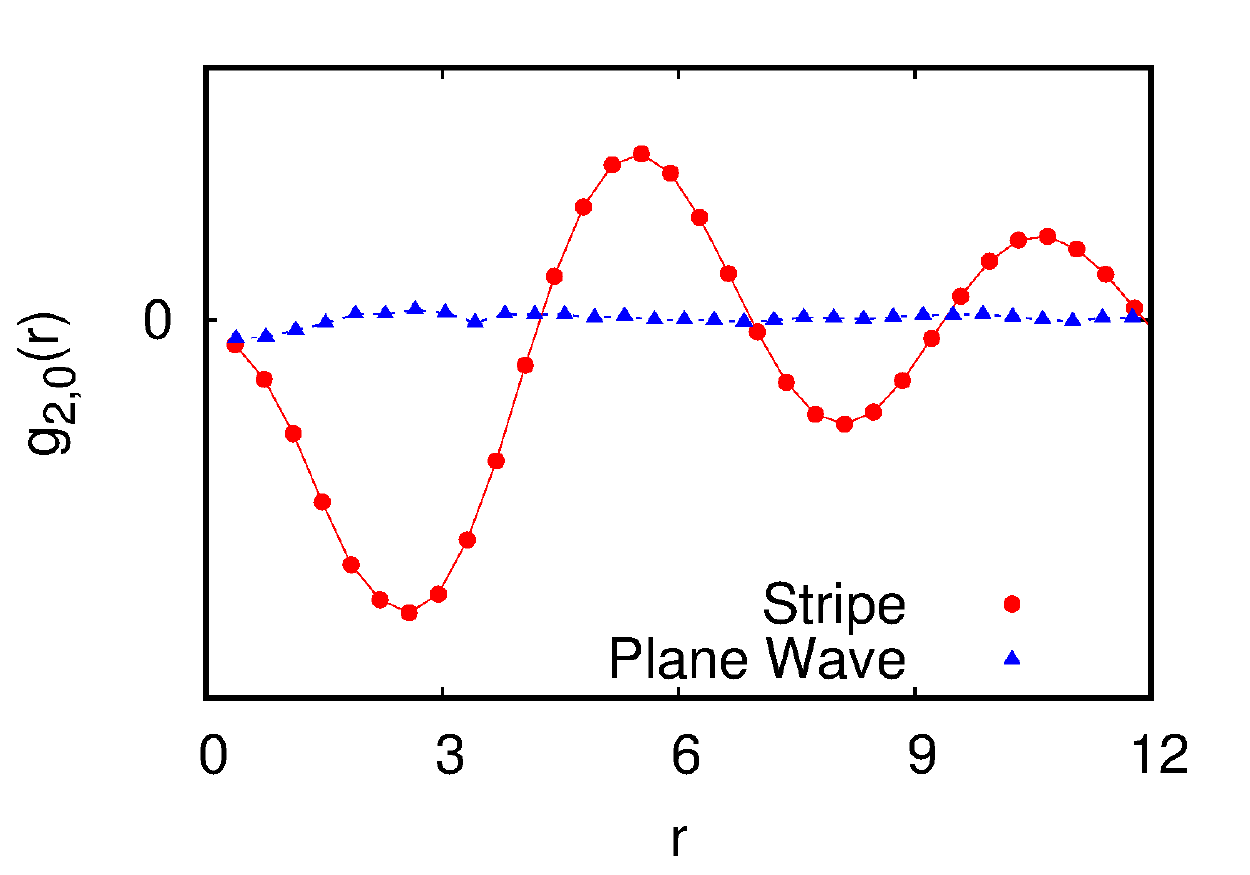}
\caption{Leading correction to the isotropic contribution to the pair
  distribution function in the ($+1,+1$) channel, corresponding to $l=2,m=0$, for the SS
  interaction.}
\label{fig_gr}
\end{figure}  

\section{\label{sec:phase_diagram}PHASE DIAGRAM AND OTHER OBSERVABLES }

The phase diagram of the SOC system is reported in Fig.~\ref{fig_diagrams}
for fixed density and varying scattering length.
The upper and middle plots correspond to the DTDMC results for the SS and 
LJ interactions, respectively, while the lower plot shows the 
mean-field phase diagram. The points indicate the computed transition lines 
between the different phases.
Errorbars in the DTDMC results account for the 
statistical variance of the energy estimations. 
Looking at the DTDMC phase diagrams it can be seen that, as the two-body 
scattering length increases, the value of the reduced Raman coupling
at which the plane wave-stripe phase transition
takes place, also increases.
Remarkably, this effect is absent at the mean-field level, and is also 
robust with respect to the interaction employed. 
Based on this, we conclude that the 
enhancement
of the stripe phase in the DTDMC phase diagrams is produced by the increase of 
inter-atomic correlations.
This enhancement takes place because the DTDMC correction to the energy of the stripe ($\Delta E_{\text{DMC,S}}$) and plane wave ($\Delta E_{\text{DMC,PW}}$) phases and the energy difference between these phases at the
mean-field level ($\Delta E_{\text{MF}}$) fulfill $ \abs{ \frac{ \Delta E_{\text{DMC,S}} }{ \Delta E_{\text{MF}} } } \simeq \abs{ \frac{ \Delta E_{\text{DMC,PW}} }{ \Delta E_{\text{MF}} } } \simeq 1$ over a wide region of the phase diagram.
The stripe phase is favored over the plane wave phase
in the DTDMC diagram because of the different polarization between
phases: while the stripe phase is always unpolarized, the plane wave
phase has non-zero polarization. Since the potentials employed in this
work are less repulsive in the $(+1,-1)$ and $(-1,+1)$ channels in
accordance to the experiment of Ref.~\cite{ketterle}, the beyond
mean-field corrections favor an unpolarized state over a polarized
one. In this sense, the DTDMC
corrections drastically determine the transition line. In contrast, the single minimum region of the diagram is only slightly changed by DTDMC with
respect to the mean-field prediction. This is because the energy gap
in mean-field between this phase and the stripe and plane wave phases is larger in absolute value than the DTDMC corrections
over the majority of the phase diagram.

In mean-field,
the stripe-plane wave and the stripe-single minimum 
transitions are of first order, while the plane wave-single 
minimum transition is of second order~\cite{review,stringari_2}.
This is directly reflected in the value of the momentum
that minimizes the energy in each phase at the mean-field level: while
there is a discontinuity in this parameter between the stripe and the
other two phases at the transition, the optimal momentum changes
continuously from the plane wave to the single minimum
phases~\cite{review,stringari_2}. We believe that the inclusion of
correlations in the DTDMC calculation does not change the nature of
any of these phase transitions.

\begin{figure}[b]
\centering
\includegraphics[width=0.85\linewidth]{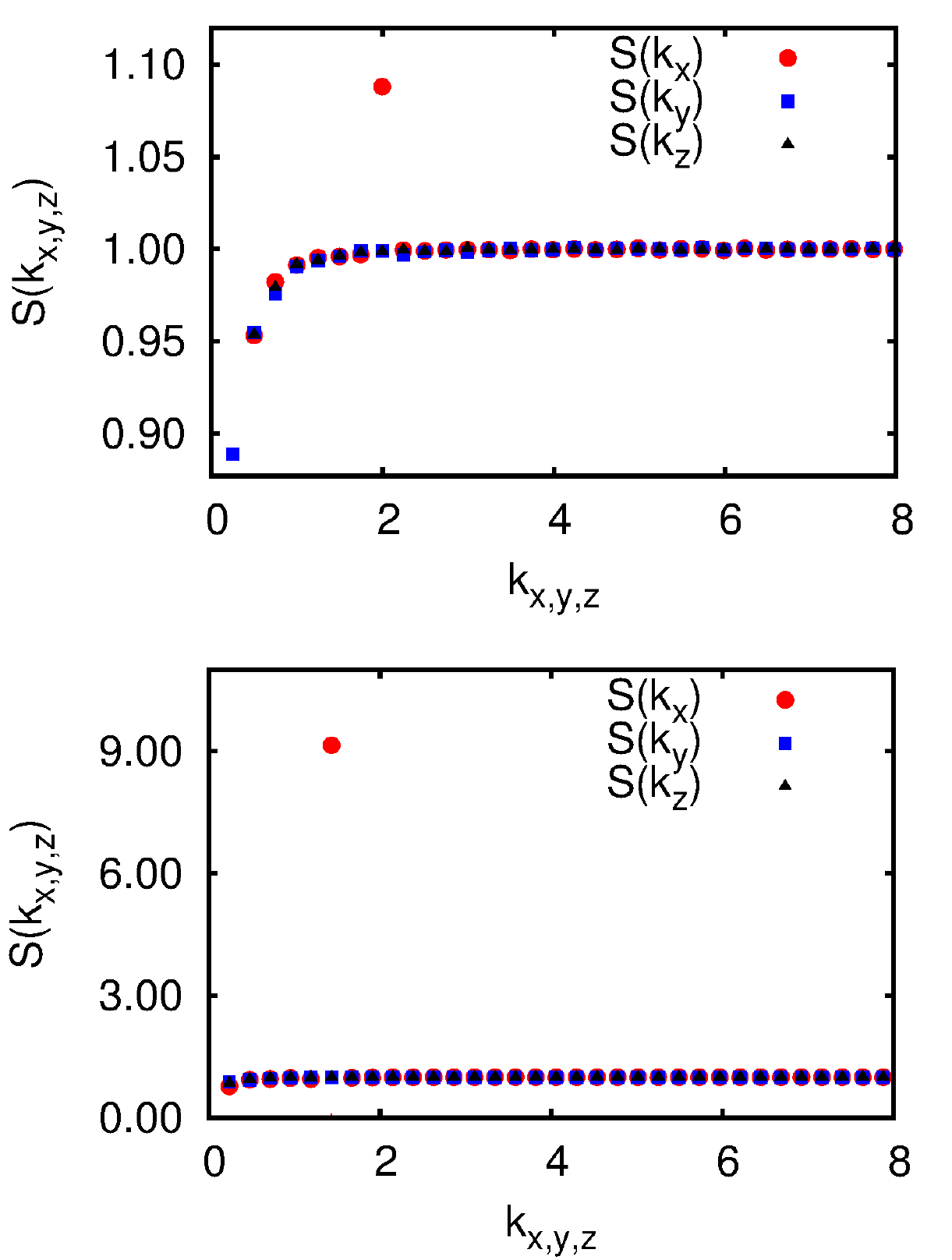}
\caption{Static structure factors for the SS interaction, for two
  different points with $n a^3 = 5 \!\cdot\!10^{-5}$ and $\gamma =
  0.904$, both corresponding to the stripe phase. 
  The upper and lower panels are for $\Omega = 0.3131$ and $\Omega =
  2.8$, respectively.}
\label{fig_superfluid_stripe}
\end{figure}

The presence of inter-atomic correlations can be seen in the pair
distribution function, $g(\vec{r}_i-\vec{r}_j)$, which yields the
probability of finding two particles with relative position vector
$\vec{r}_i-\vec{r}_j$. For an
isotropic system, $g(\vec{r})$ depends on $\abs{\vec r}$, while for a non-isotropic system, as
it is the case of the stripe phase,
an expansion in partial waves of the form
$g(\vec{r}_i-\vec{r}_j) = \sum_{l,m} g_{l,m}(r_{ij}) Y_l^m(\theta, \phi)$
yields non-zero contributions for $l>0$. Notice that, in this expression,
$\theta$ is the angle formed by $\vec r$ and the $x$-axis, and stripes
are formed along planes perpendicular to that direction.
In Figure~\ref{fig_gr}, we show the leading correction to the
isotropic mode, for two points in the phase diagrams corresponding to
the stripe and plane wave phases. Only the ($+1,+1$) component is
reported since results for the rest of the two-body channels are
analogous.  The Figure depicts the $l=2, m=0$ modes for the SS
interaction. It should be pointed out that, for the specific type of
interactions used in this work, only the $m=0$ contributions survive.
As one can see, $g_{2,0}(r_{ij})$ is zero in the
  plane wave phase, while it yields a non-vanishing contribution in
  the stripe phase. This reflects the different spatial symmetries
  associated to each phase~\cite{stringari}.  This quantity also
  vanishes for the single minimum phase.
Very similar results
hold for the LJ interaction.

Since in the stripe phase the $x$-axis is transverse to the stripe
planes, the static structure factor along the $x$-direction,
$S(k_x)$, develops a peak at a momentum proportional to the
inverse of the characteristic distance separating the stripes,
a feature also present at the Bogoliubov
  level~\cite{stringari_stripes}.
We show in the upper panel of
Fig.~\ref{fig_superfluid_stripe} the static structure factor $S(\vec
k)$ for conditions similar to the experiment of Ref.~\cite{ketterle}
($\gamma=0.904$, $\Omega=0.3131$ and $na^3 = 5\cdot 10^{-5}$).  The
lower panel shows the same quantities for $\Omega=2.8$, where the
stripe modulation is much more important due to the higher value of
$\Omega$ compared to $E_0$. In accordance to that experiment, where
reduced Raman coupling values lay in the interval $\Omega_{\text{exp}}
\in [0,\text{ }0.4]$, we recover the stripe phase as the lowest energy
state.
The periodicity of the stripes has been
  quantitatively characterized before both in the mean-field
  regime~\cite{stringari,stringari_stripes} and in
  experiments~\cite{ketterle}.
The static structure factor does not show any peak in
  the plane wave and single minimum phases, a consequence of the lack
  of density modulations in these phases~\cite{stringari}.

\begin{figure}[t]
\centering
\includegraphics[width=0.95\linewidth]{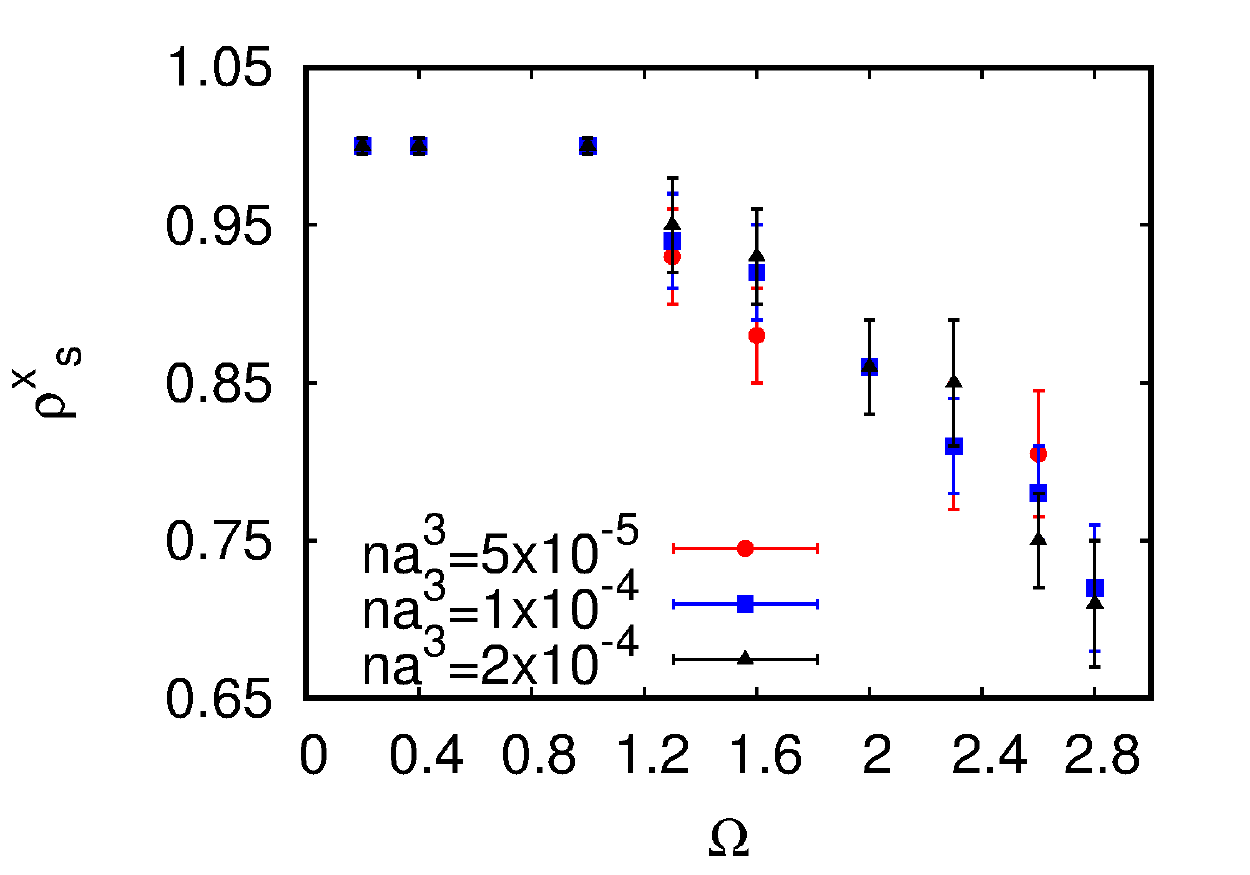}
\caption{Superfluid fraction for the transverse direction to the stripe
planes as a function of the reduced Raman coupling for $n a^3 = 5 \!\cdot\! 10^{-5}$, 
$\gamma = 0.904$.} 
\label{fig_superfluid_vs_omega_na3}
\end{figure}

Finally, we characterize the superfluidity of the system in the stripe
phase, where other systems have shown a non-trivial dependence along
different directions~\cite{Bombin}.  In order to recreate the
conditions of contrast and diluteness of Ref.~\cite{ketterle}, we set
$\gamma = 0.904$ and use gas parameters spanning the range $n a^3 \in
[5 \!\cdot\!  10^{-5}, \text{ } 2 \!\cdot\! 10^{-4}]$.  We measure the
superfluid density using the zero-temperature limit of the winding
number estimator~\cite{zhang95}, which is extracted from the mean
squared displacement of the center of mass of the particles during
imaginary time evolution.
We show in Fig.~\ref{fig_superfluid_vs_omega_na3} results for the
superfluid fraction $\rho_s^x$ in the stripe phase along the $x$
direction, obtained from the generalization of the expression reported
in Ref.~\cite{Bombin}, as a function of $\Omega$, and for three different
values of the gas parameter, $n a^3 = 5 \cdot\!  10^{-5}, 1\!\cdot\!
10^{-4}$, and $2 \!\cdot\!  10^{-4}$.
The reported values are close to those
  present in Ref.~\cite{superf_1}, obtained at the mean-field level
  using the twisted phase method.
We see from the plot that the main parameter governing changes in
$\rho_s^x$ is $\Omega$, while little dependence on the specific value
of the gas parameter is found.
As $\Omega$ increases, the system becomes less
superfluid in the $x$ direction. This is a direct consequence of the
fact that the amplitude of the density modulation increases with
$\Omega$, 
as already seen in mean-field theory.
For large values of $\Omega$, exchanges of particles between different stripe
planes are less favored, and thus localization along the $x$ axis is
enhanced.
In the other two directions, parallel to the stripe planes, the system
remains fully superfluid ($\rho_s^y=\rho_s^z=1$).
Notice also that, for the values of $\Omega$
employed in the experiment of Ref.~\cite{ketterle}, the superfluid
fraction $\rho_s^x$ equals one. This, together with the periodic
density modulations in the static structure factor reported in
Fig.~\ref{fig_superfluid_stripe}, yields a quantitative indication of
simultaneous spatial periodicity and superfluidity in the
system.

The superfluid fraction for the plane wave and single
  minimum phases has been obtained at the mean-field level using the
  phase twist method~\cite{superf_1} and in the Bogoliubov model
  through the evaluation of the transverse current
  operator~\cite{superf_2}. In this case, the superfluidity along the
  $y$ and $z$ axes in these phases is equal to unity, while $\rho_x^s$
  shows a dependence on the Raman coupling. We recover these results
  with DTDMC for the gas parameters mentioned previously by using the
  expression for the normal density of Ref.~\cite{superf_2}, replacing
  the mean-field value of $\langle \sigma_x \rangle$ by the one
  provided by DTDMC.

\section{\label{sec:conclusions}CONCLUSIONS }

In summary, we have shown, using the DTDMC method~\cite{sanchez,
  mitas}, that by increasing the strength of inter-atomic correlations
in a system under Raman SOC, the region of the phase diagram covered
by the stripe phase is enlarged in comparison to the prediction of
mean-field theory.
We have shown that this effect holds for different two-body model
interactions (soft spheres and Lennard-Jones), which provide very
similar results.  The breaking of continuous translational symmetry in
the stripe phase has been characterized by the presence of a Bragg peak in
the static structure factor, and by a nonzero contribution to partial
waves other than the $l=0$ to the pair distribution function.
We have also performed DTDMC calculations in the same conditions of
interaction contrast and reduced Raman coupling of the experiments of
Ref.~\cite{ketterle}. Our results confirm the observed stripes as the
most energetically favorable state and quantitatively show the
supersolid behavior of the stripes.
We have also studied the superfluid fraction of the stripe phase as a
function of the reduced Raman coupling and the gas parameter by changing the
scattering length of the inter-atomic interaction. We have
shown that superfluidity in the stripe phase decreases mainly as a
function of the reduced Raman coupling, with little dependence on the gas
parameter in the range analyzed.  We hope that our work can encourage
possible experimental studies of Raman SOC systems near the transition lines
between the stripe and the plane wave
phases, since the effects of correlations beyond the mean-field
approximation can be seen even at relatively low gas parameter values
like $n a^3 = 10^{-4}$.

\section{\label{sec:ACKNOWLEDGMENTS}ACKNOWLEDGMENTS }

We acknowledge interesting discussions with S. Stringari on the superfluidity of the different SOC phases.
This work has been supported by the MINECO (Spain) Grant No. 
FIS2017-84114-C2-1-P. J. S\'anchez-Baena also acknowledges the FPU fellowship with reference FPU15/01805 from MECD (Spain).

\end{document}